\documentclass[10pt,letterpaper]{article}

\usepackage[utf8]{inputenc}

\usepackage{cite}

\usepackage{nameref,hyperref}

\usepackage{microtype}
\DisableLigatures[f]{encoding = *, family = * }

\usepackage{epstopdf}

\usepackage{color}

\usepackage{booktabs} 
\usepackage{graphicx}  
\usepackage{caption}
\usepackage{subcaption}
\usepackage{amsmath}
\usepackage{amsfonts}
\usepackage{multirow}

\newcommand{\ra}[1]{\renewcommand{\arraystretch}{#1}}

\definecolor{Gray}{gray}{.25}

\usepackage{graphicx}

\usepackage{amsthm}

\usepackage{wrapfig}
\usepackage[pscoord]{eso-pic}

\begin{document}
\vspace*{0.35in}

\begin{flushleft}
{\Large
\textbf\newline{A Content-Based Approach to Email Triage Action Prediction: Exploration and Evaluation}
}
\newline
\\
Sudipto Mukherjee\textsuperscript{1},
Ke Jiang\textsuperscript{2},
\\
\bigskip
\bf{1} University of Washington, Seattle, WA, United States.
\\
\bf{2} Microsoft Corporation, Bellevue, WA, United States.
\\
\bigskip
 sudipm@uw.edu, kejian@microsoft.com

\end{flushleft}

\section*{Abstract}

Email has remained a principal form of communication among people, both in enterprise and social settings. With a deluge of emails crowding our mailboxes daily, there is a dire need of smart email systems that can recover important emails and make personalized recommendations. In this work, we study the problem of predicting user triage actions to incoming emails where we take the reply prediction as a working example. Different from existing methods, we formulate the triage action prediction as a recommendation problem and focus on the content-based approach, where 
the users are represented using the content of current and past emails. We also introduce additional similarity features to further explore the affinities between users and emails. Experiments on the publicly available Avocado email collection demonstrate the advantages of our proposed recommendation framework and our method is able to achieve better performance compared to the state-of-the-art deep recommendation methods. More importantly, we provide valuable insight into the effectiveness of different textual and user representations and show that traditional bag-of-words approaches, with the help from the similarity features, compete favorably with the more advanced neural embedding methods.

\section{Introduction}
\label{sec:intro}
\noindent  
The advancement of artificial intelligence has brought forth many smart client systems aimed at improving user experience. Email service providers, such as Microsoft Outlook\footnote{https://outlook.live.com/} and Gmail\footnote{https://www.google.com/gmail/}, have started to incorporate intelligent features into their products ranging from organizing the received emails into different folders based on the inferred priority to reminding users to reply to an email which has not been attended to for long. Thus, this work aligns nicely with the \textit{business} relevance of email service providers such as Focused Inbox in Microsoft Outlook and the Nudge application in Gmail. Since millions of people all over the world send and receive emails daily, our work aims to significantly ease users' communication load and increase the triage productivity. In these applications, one of the core components is the email triage action prediction. 

There are two flavors to this problem of email triage action prediction. We can view it from the perspective of a sender where an email is sent and the system predicts the probability of receiving response from \textit{any} of the recipients \cite{yang} (denote \textit{Problem S}). The other is to view it from the recipient's end \cite{gmail}. In this scenario, we want to predict for each recipient of an email whether he/she will take some action on it, which is clearly more challenging (denote \textit{Problem R}). Previous studies have investigated different categories of features for both problems, but mostly focused on social and people affinity features \cite{gmail,yang}. However, existing methods often ignore the email content and can not fully capture users' unique action patterns.

In this work, we study the more challenging problem of email triage action prediction from the recipient's perspective (\textit{Problem R}), aiming to fill this gap and focus primarily on the email content. Intuitively, the content of the incoming email alone is insufficient for prediction since the same email could be received by multiple people and some may respond while others may not. Even for the much simpler \textit{Problem S}, \cite{yang} could derive little predictive power from the email content with naive bag-of-words tokens. This calls for some personalization or appropriate representation of the email and recipients, which can be naturally formulated as a recommendation problem. Traditional recommendation approaches, which model the user using a one-hot encoding or their learned embedding variants \cite{gmail,ncf}, can not automatically extend to new users or dynamically adapt to existing users' changing behaviors. Therefore, it has to be re-trained frequently \cite{gmail} and hence is not scalable.

Our email triage action prediction framework follows the content-based approach \cite{content}. It models the recipients using their historically received emails and takes the temporal order into consideration. The input consists of the incoming email and the recipient's Inbox history. The output is the probability that the recipient would perform certain action on the email. We systematically investigate the potency of different content representations ranging from traditional bag-of-words to more advanced neural embedding representations, and various Inbox history selections considering both ``supervised" and ``unsupervised" ways. In addition, to further explore the affinities between users and emails, we also introduce the similarity features which will facilitate the prediction.
Thorough experiments on the publicly available Avocado email collection show the benefits of our proposed recommendation framework and provide valuable insight into the effectiveness of different representations. 

Note that, although we focus on the email content, other features like people affinity and meta-data studied in existing methods can be easily incorporated as additional features in our framework. 
However, extensive study on the impact of various content-based user representations has not been explored in the literature and thus the main focus of our work. In addition, our recommendation framework is general and can be applied to any online recommendation problem. Our findings provide interesting directions to a practitioner in the field, especially since email data is inherently different from Tweets or News highlights. Emails have huge diversity in their content and length; so incorporating contrastive features from traditional models alongside advanced neural architectures is deemed useful from our experiments.  

\textbf{Contributions}. The main contributions of our work are summarized as follows.
\begin{itemize}
\item We propose the recommendation framework of content-based approach for email triage action prediction problems. To the best of our knowledge, this is the first work that systematically studies this problem considering a variety of content-based email and user representations.

\item We illustrate the effectiveness of representing recipients using their historically received emails. Class-specific recipient representation delivers the best performance among all representations considered, and shows improvement over the widely used positive-only representation \cite{blei,dkn,interest}. 

\item We show that the similarity features introduced between the incoming and past emails are critical to deliver additional performance lift. Indeed, we are able to outperform recently proposed deep recommendation methods \cite{ncf,dkn,interest,vartak} by incorporating them. More importantly, we find that traditional bag-of-words approaches, with the help of the similarity features, compete favorably with more advanced neural embedding representations.

\item Finally, we show that bag-of-words and neural embedding representations contain complementary information and are able to obtain additional improvement with a simple ensemble model.

\end{itemize}

The rest of the paper is organized as follows. Section \ref{sec:model} introduces the proposed framework and its various manifestations. The experimental results are provided in Section \ref{sec:experiment}. Section \ref{sec:relatedwork} discusses related works in this area and differentiates our approach from existing literature. We conclude the paper with some future directions in Section \ref{sec:conclusion}.

\section{The Framework}
\label{sec:model}
Predicting user's triage actions to incoming emails, such as reply, flag and delete, is of great importance and would largely help people prioritize their mailboxes more effectively. In this section, we take the reply prediction as a working example to formulate the recommendation framework. The task is to build a binary classification model to estimate if the user would reply to a particular incoming email. 

Specifically, for an incoming email $e_i$ to a recipient $u_j$, the outcome $r_{ij}\in \{0,1\}$ indicates if the recipient would reply to this email. Our goal is to predict the probability that $u_j$ will reply to $e_i$:
\begin{equation}
	\mathbb{P}(r_{ij} = 1 \,|\, e_i, u_j) = f(e_i, u_j).
\end{equation}
We focus on the content-based approach where the email $e_i$ is represented by its textual content and recipient $u_j$ is represented using the content of her Inbox history. In this way, the recipient representations are not constrained by the patterns limited in the training set and will be automatically updated as new emails arrive, thus naturally adapting to changing reply patterns. 

\begin{figure}[h]
\centering
\includegraphics[width=\textwidth]{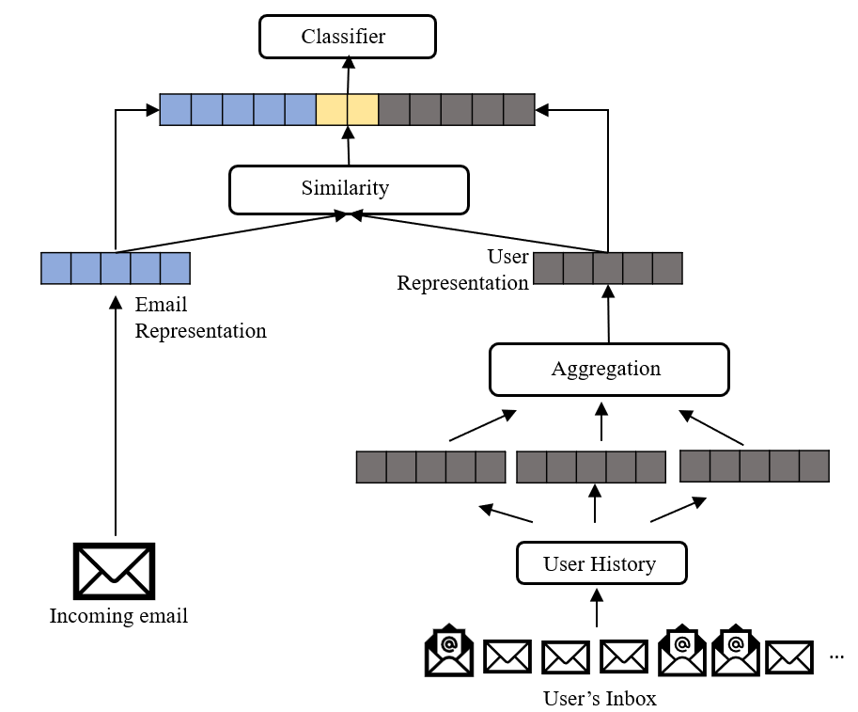}
\caption{Framework of the proposed content-based email triage action prediction.}
\label{fig:framework}
\end{figure}

The proposed framework is illustrated in Figure \ref{fig:framework}. It takes the incoming email and a set of the recipient's email history as input. For each email, a textual feature extractor is used to process its content and generate a representation vector. We consider different extractors, from traditional to more advanced deep learning methods (Section \ref{sec:email}). To get the final user representation, we use an aggregation module to combine different emails into one summarization. Both global transformation and attention-based methods \cite{attention} are considered (Section \ref{sec:user}). Unlike previous methods which directly concatenate the email and user embeddings together and feed into a deep network \cite{ncf,dkn,interest}, we apply an additional similarity module (Section \ref{sec:simfeat}) on top of the email and user embeddings to extract further affinity relationships between them. Finally, the email representation, user representation, and the similarity features are concatenated together and fed into a classifier to calculate the probability that the recipient will reply to the incoming email. In this framework, the representation extractor and the classifier are both shared among all the users which will enhance knowledge sharing among different users.

\subsection{Email Representations}
\label{sec:email}
For the email representations, we turn to its textual contents and consider two representatives from both sides of the spectrum of text feature representations. 
\begin{itemize}
\item The first is the sparse and high-dimensional bag-of-words representation and its term-frequency based variants \cite{bow}, which is widely used in natural language applications. We use the tf-idf variant in our experiments where each email is represented as a vector whose dimension is the same as the vocabulary size and the values are the tf-idf weights of the corresponding terms extracted from the email.
\item The other one is the neural word embeddings learned from some corpus, unsupervised \cite{word2vec,glove,fasttext} or supervised \cite{kim}, which are dense and low-dimensional. Since the embedding is at the word level, we either use the simple average of all the word embeddings for the corresponding tokens in the email, or use the convolutional neural network \cite{kim} trained on top of the word embeddings from the supervised signal to get the email representation. We choose convolutional neural network in our experiments since it has empirically been shown to be competitive and can be more effective for lengthy emails compared to recurrent neural networks.
\end{itemize}

\subsection{User Representations}
\label{sec:user}
We represent each user in terms of the content of their inbox history, i.e., the set of emails that they have received. On one hand, it still provides personalization based on the distinct contents from different received histories. On the other hand, it can accommodate new users automatically once they start receiving emails. We select up to $h$ \textit{most recently} received emails that were received before the new incoming one for representing users. This also provide a natural temporal adaptation to the users' changing behaviors. 

\begin{figure}[h]
\centering
\includegraphics[width=0.5\textwidth]{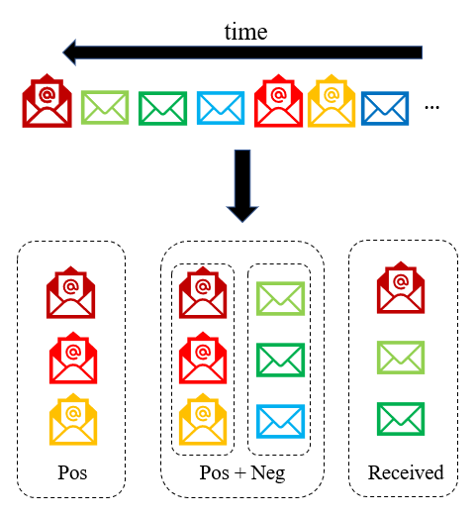}
\caption{Illustration of different user history selections considered in this work. Here, $h=3$ and opened emails represent replied ones while unopened represents non-replied ones. Best viewed in color.}
\label{fig:user}
\end{figure}

We consider the following three different categories of history selections, which is illustrated in Figure \ref{fig:user}:
\begin{itemize}

\item The positive history (\texttt{Pos}): we only consider the most recently received $h$ emails that have been replied to before receiving the new email. This provides information regarding historically what kinds of content would get reply from the user.

\item The positive and negative history (\texttt{Pos+Neg}): here, we consider two sets of histories. As shown in Figure \ref{fig:user}, in addition to the positive history introduced above, we consider its negative counterpart which is the most recently received $h$ emails\footnote{We consider the same number of positive and negative historical emails to create a balanced history.} without reply as well. This not only provides labeled information from both classes (replied/non-replied) but also is essential to obtain the crucial contrastive term that we will explore below in Section \ref{sec:simfeat}.

\item The received history (\texttt{Received}): we also consider directly using the received history without attending to the actions performed upon. This only serves as a user profile representation since there is no way to tell what have been replied and what have not. As far as we know, this kind of user history has not been explored in the literature. Nevertheless, we find that this representation still performs competitively with the widely used positive history.

\end{itemize}

\noindent
\textbf{Aggregation}. Given the selected user history $\{e_1^j, \ldots, e_h^j\}$, we summarize them into one representation vector with the same dimension as the email counterpart. That is, the resulting user representation can be written as:
\begin{equation}
	\mathcal{G}(u_j) = \mathcal{G}(\mathcal{F}(e_1^j),\ldots, \mathcal{F}(e_h^j)) = \sum_{t=1}^{h}\alpha_{t}\mathcal{F}(e_{t}^{j}),
\label{eqn:aggregate}
\end{equation}
where $\mathcal{F}$ is the email feature extractor, and $\{\alpha_t\}_{t=1}^{h}$ are the aggregation weights. For the \texttt{Pos+Neg} history, we compute two representation vectors where one comes from the positive history and the other from the negative history respectively. 

The aggregation weights $\{\alpha_t\}_{t=1}^{h}$, in a global case, can simply be all $1/h$ or trainable parameters which are shared among users. However, such global aggregation approaches ignore the varying relationships between the incoming email and the history, where different historical emails may have different impacts on the incoming email (when considering if the user will reply to it). To get an email-specific user representation, we resort to an attention module \cite{attention} to model the different impacts of the historical emails on the incoming one. Specifically, the weights $\{\alpha_t\}$ are dependent on $e_i$ and can be written as:
\begin{equation}
	\alpha_{t}^{i} = \frac{\exp(\gamma\cdot \text{score}(\mathcal{F}(e_i), \mathcal{F}(e_t^j)))}{\sum_{t'}\exp(\gamma \cdot\text{score}(\mathcal{F}(e_i), \mathcal{F}(e_{t'}^j)))},
\label{eqn:attention}
\end{equation}
where $\gamma$ is a tuning parameter and the score function computes the semantic relationship between the two email representations. We consider the following two score functions \cite{attention}: 
\begin{equation}
 \text{score} = \mathcal{F}(e_i)^\top \mathcal{F}(e_{t}^{j}) \quad\text{(dot)}, 
\label{eqn:dot}
\end{equation}
and 
\begin{equation}
\text{score} = v^\top \text{relu}(W[\mathcal{F}(e_i); \mathcal{F}(e_t^j)])\quad \text{(concat)},
\label{eqn:concat}
\end{equation}
where $\text{relu}(x) = \max(x,0)$, $v$ and $W$ are trainable parameters.

In the following sections, \texttt{Pos}, \texttt{Pos+Neg} and \texttt{Received} will refer to the corresponding aggregated user representations.

\subsection{Similarity between Email and User}
\label{sec:simfeat}
We consider the global similarities between the email and user representations as additional features to the final classifier. These are particularly important especially for the \texttt{Pos} and \texttt{Pos+Neg} representations where they serve more like class ``prototypes" \cite{prototype} and the similarities themselves can provide enough predictive power already. In addition, these similarities could also free the classifier to attend more to the local details and interaction patterns instead of the global resemblance.

We use the simple inner product between the email and user representations in our experiments. For incoming email $e_i$ and the user representation $\mathcal{G}(u_j)$, the similarity is computed as:
\begin{equation}
	\mathcal{S}(\mathcal{F}(e_i), \mathcal{G}(u_j)) = \mathcal{F}(e_i)^\top\mathcal{G}(u_j).
	\label{eqn:sim}
\end{equation}
For the \texttt{Pos} and \texttt{Received} user representations, we only compute the similarity feature using \eqref{eqn:sim}. For the \texttt{Pos+Neg} user representation, the similarity features consist of three parts: the similarity with the positive representation , the similarity with the negative representation, and the difference between those two similarities. Despite seeming redundant considering the power of the nonlinear classifiers, we will demonstrate the advantages of including the similarity features even for deep neural networks in the experiments section.

\section{Experiments}
\label{sec:experiment}
In this section, we present detailed quantitative analysis regarding different textual representations along with various user embeddings discussed in Section \ref{sec:model}, and also show the performance comparison with existing deep recommendation methods.

\subsection{Dataset}
The dataset used in our study is the Avocado research email collection \cite{avocado}. It consists of enterprise emails collected from $279$ accounts of a defunct information technology company named ``Avocado". The entire dataset consists of $938,035$ emails including duplicates\footnote{Spam emails were not included in the dataset.}. We restrict ourselves to the emails received from June 1st 2000 to May 31st 2001 and generate the ground truth for the reply action using the existing ``reply\_to" field. Duplicated emails are discarded. Emails with missing sender name or sent date are eliminated as well. We also eliminate emails where the sender is the only recipient and exclude recipients who have not replied to any email during the entire time period considered. They are either some group aliases or recipients outside of the company for which the dataset has no reply information. 

We focus on the reply prediction for the first email in a conversation thread for each recipient individually. Therefore each email id in the dataset will give rise to multiple received emails according to the number of recipients, as we consider the reply prediction from the recipient's perspective. The resulting email collection contains $429,084$ emails received by $268$ unique recipients. We split the collection into training/validation/testing sets respecting the temporal order. To be more precise, we use emails received from June 1st 2000 to Jan 31st 2001 for training, emails received from Feb 1st 2001 to Feb 28th 2001 for validation and those from March 1st 2001 to May 31st 2001 for testing. In this way, we ensure that no future information is used to train the model. The basic statistics of the partitions and the distribution of the reply action are shown in Table \ref{tab:summary}.

\begin{table}[h]
\centering \ra{1.2}
\begin{tabular}{ @{} rrrr@{} } 
 \toprule
  & \textbf{Training} & \textbf{Validation} & \textbf{Test}\\ \midrule
\#Recipients & 246 & 223 & 230\\ 
\#Incoming emails & 244,532 & 45,051 & 139,501 \\ 
\#Emails w/ reply & 18,814  & 4,673  & 13,691 \\ 
Positive ratio & 7.7\% & 10.4\% & 9.8\%\\
\bottomrule
\end{tabular}
\caption{Summary statistics of the resulting dataset.}
\label{tab:summary}
\end{table}

\subsection{Experimental Setup}
\textbf{Text Processing}. We first clean the email text by removing all the words after ``original message'', and then build the vocabulary of the most frequent $10$k word $N$-grams with $N$ up to $3$ from the training and validation sets only, after removing stop words that appear in more than $95\%$ of the emails. Named entities and email aliases are removed as well\footnote{Including the named entities would give rise to a vocabulary, and consequently to a feature set, prone to overfitting the user interactions in training data. They offer limited generalization to unseen users with the learned people relationships in a closed universe. It can not adapt to changing interactions with time either. Thus, we need to ensure that named entities are not part of the content features in our model. On the other hand, the people affinity features can be directly captured \cite{gmail,yang} and incorporated in our framework easily.}. 

\vspace{2pt}
\noindent
\textbf{Email Representations}. We consider the following three variants: 
(i) the tf-idf weighted bag-of-words representation (TFIDF); 
(ii) the average of the fasttext \cite{fasttext} word embeddings of the tokens in the email (Embed), where the embedding dimension $100$ is used; 
(iii) the representation learned using convolutional neural network on top of the fasttext word embeddings (CNN). 
The email content is restricted to the first $150$ tokens\footnote{Note this is considerably longer than the length of the news titles or tweets previously studied in the recommendation literature.} with zero padding. Here, we use CNN to indicate the representation learning part only. The fasttext embeddings in (ii) and (iii) are learned from solely the combined training and validation email contents without using any test email. We use the same CNN architecture as in \cite{kim} to obtain the representation for an email, which is shared between current and past received emails. 

\vspace{2pt}
\noindent
\textbf{User Representations}. For user representations, the following settings are used if not stated otherwise. The length of history considered is up to $10$. For the aggregation, simple average ($\alpha_t = 1/h$ in \eqref{eqn:aggregate}) is applied for both TFIDF and Embed representations, while dot-based attention weights \eqref{eqn:attention} and \eqref{eqn:dot} are used for the CNN representations.

\vspace{2pt}
\noindent
\textbf{Classifiers and Training Details}. We experiment with various binary classifiers including logistic regression (LR), gradient boosted decision trees (GBDT), and multilayer perceptron (MLP) with the weighted log-loss (positive labeled data points are weighted to counter the acute class imbalance in the problem) .  Similar to \cite{vartak}, we include the recipient reply rate as an additional feature to the classifier. We use LightGBM \cite{lightgbm} to learn the boosted decision trees. The LR, CNN and MLP models are implemented in TensorFlow \cite{tensorflow} and optimized using Adam \cite{adam}. Random search \cite{randomsearch} is employed to tune the hyper-parameters (details can be found in the Appendix). For deep models, we perform five runs with the best parameters found based on the validation performance and report the averaged test result with standard deviation. We find that the standard deviation in general is smaller than $0.003$.

\vspace{2pt}
\noindent
\textbf{Evaluation}. Since the dataset is highly imbalanced, we use the area under the receiver operating characteristic curve (AUROC) for evaluation, which is a ranking metric insensitive to class imbalance. Higher AUROC values indicate better performance, while random guessing gives $0.5$ value.

In the following, we will use the ``Content-Classifier" notation to indicate the model used, where ``Content" is the email representation considered and ``Classifier" is the binary classifier employed. For example, ``TFIDF-GBDT" indicates the GBDT classifier performed on top of the TFIDF content representations, while ``CNN-MLP" stands for the MLP classifier performed on top of the CNN content representations.

\subsection{Baselines}
We consider the following state-of-the-art deep recommendation methods as baselines in our experiments:
\begin{itemize}
\item \textbf{Neural collaborative filtering}\cite{ncf} is a deep generalization of the matrix factorization method. It leverages a multi-layer perceptron to learn the user-item interaction function. The user and item representations are both learned through explicit embedding on top of the one-hot indicator vectors with learnable embedding matrices. We use an adapted version where only the user representation is learned through explicit embedding since the incoming emails are always new and the ID-based method can not be applied for emails.
\item \textbf{Positive content-based recommendation} \cite{dkn,interest} is a family of methods where the user representation is generated by using the attention module to aggregate the contents of the positive history. This is similar to the \texttt{Pos} user presentation considered in our work.
\item \textbf{Meta-learning based item recommendation} \cite{vartak} is the start-of-the-art adaptation based recommendation method to address the item cold-start problem. It uses both positive and negative user histories to generate the user representations and proposes two adaptation classifiers which take advantage of those: linear classifier with weight adaptation (LWA) and nonlinear classifier with bias adaptation (NLBA).
\end{itemize} 

\begin{table}[h]
\centering \ra{1.2}
\begin{tabular}{@{}rrrrr@{}}\toprule
\multirow{2}{*}{\textbf{Content}} & \multirow{2}{*}{\textbf{Classifier}} & \multicolumn{3}{c}{\textbf{User Representation}} \\ \cmidrule{3-5}
 & & \texttt{Received} & \texttt{Pos} & \texttt{Pos+Neg} \\ \midrule
\multirow{2}{*}{TFIDF} & LR & 0.7455 & 0.7143 & 0.7628 \\
 & GBDT & \textbf{0.7622} & 0.7248 & \textbf{0.7773} \\  \midrule
\multirow{2}{*}{Embed} & LR & 0.7213 & 0.7108 & 0.7374 \\ 
  & GBDT & 0.7409 & 0.7236 & 0.7531 \\ \midrule
\multirow{2}{*}{CNN} & \multirow{2}{*}{MLP} & 0.7592 & \textbf{0.7457} & 0.7764 \\
 & & (0.0013) & (0.0029) & (0.0007) \\ \bottomrule
\end{tabular}
\caption{Test AUROC results of different email and user representations with history length up to $10$ using various binary classifiers. The numbers in the parentheses are the standard deviations for the CNN-MLP model. The best performance for each user representation is shown in bold.}
\label{tab:meta}
\end{table}

\subsection{Results}
In this section, we present the results of comparison among different variants of the proposed framework and the comparison with existing deep recommendation models.

\subsubsection{Comparison of Different Representations}
The test AUROC results of different variants with various email and user representations are shown in Table \ref{tab:meta}. We can see that the email content has already expressed reasonable predictive power once the user representation is incorporated. Similar to other classification problems, nonlinear classifiers give better prediction performance compared with the linear ones irrespective of the email and user representations used. We also experiment with the CNN-LR model whose averaged result is $0.7598$. It still has more than $2\%$ AUROC decrease compared to its nonlinear counterpart. This shows that the interaction between the contents from email and user representation is crucial to the reply prediction.

In terms of user representations, the \texttt{Pos} user representation performs worst among all three representations considered, despite its wide adoption in the recommendation literature \cite{blei,dkn,interest}. The performance even lags far behind the ``unsupervised" \texttt{Received} user representation which does not use any historical reply information. We can see that the best result under the \texttt{Pos} user representation, which comes from the most complicated CNN-MLP model, is only comparable with that from the simple TFIDF-LR model under the \texttt{Received} user representation. This shows that there may not be enough information due to the limited number of emails that the users have replied to. 
Combining positive and negative histories together gives performance lift and further exceeds the personalization experience provided by using the \texttt{Received} user representation. The benefit from the additional negative history is most significant for the bag-of-words representation, where the performance is improved at least $6\%$ from using the positive history only. We will show in Section \ref{sec:length} that the improvement comes from incorporating the negative history, and not due to more number of historical emails considered.

As for the textual representations, we do not observe any improvement in using the averaged word embeddings over tf-idf, irrespective of the user representation and classifier considered. This suggests that the important interactions might happen due to particular terms occurring in the email, rather than at the language semantic level. The simple average operation can have a detrimental effect, specially if lexical attributes dominate.
CNN representation on top of the word embeddings, on the other hand, detects and picks up specific local patterns through the convolution operations from the emails. Therefore, it provides considerable improvement over the Embed representation. Moreover, it delivers the best performance when only the positive history is considered with at least $2.9\%$ AUROC improvement comparing to other representations.

Finally, we observe the somewhat surprising result that the CNN representation is unable to show significant difference from the traditional bag-of-words representations when the \texttt{Pos+Neg} user representation is used, which is also the best-performing user representation in our analysis. In fact, the TFIDF-GBDT model performs competitively with the more advanced CNN-MLP method in this scenario. The reason behind this is the similarity features introduced in Section \ref{sec:simfeat} and we will demonstrate it in detail in the following section.

\subsubsection{Importance of the Similarity}
\label{sec:sim}
We demonstrate now the importance of the similarity features, especially the contrastive term introduced for the \texttt{Pos+Neg} user representation, which are not only the basis of the competitive performance from the TFIDF-GBDT model but also the reason behind the state-of-the-art performance comparing with existing deep recommendation methods shown in Secton \ref{sec:compexist}.

\begin{table}[h]
\centering \ra{1.2}
\begin{tabular}{@{}rrr@{}}\toprule
\textbf{Model} & \textbf{w/o similarity} & \textbf{w/ similarity} \\ \midrule
TFIDF-GBDT & 0.7442 & 0.7773 \\
CNN-MLP & 0.7541 & 0.7764 \\ \bottomrule
\end{tabular}
\caption{Comparison of test AUROC results of models with and without similarity features where users are represented using the \texttt{Pos+Neg} representation with history length up to $10$.}
\label{tab:sim}
\end{table}

\begin{figure}
\centering
    \begin{subfigure}[b]{0.48\textwidth}
        \includegraphics[width=\textwidth]{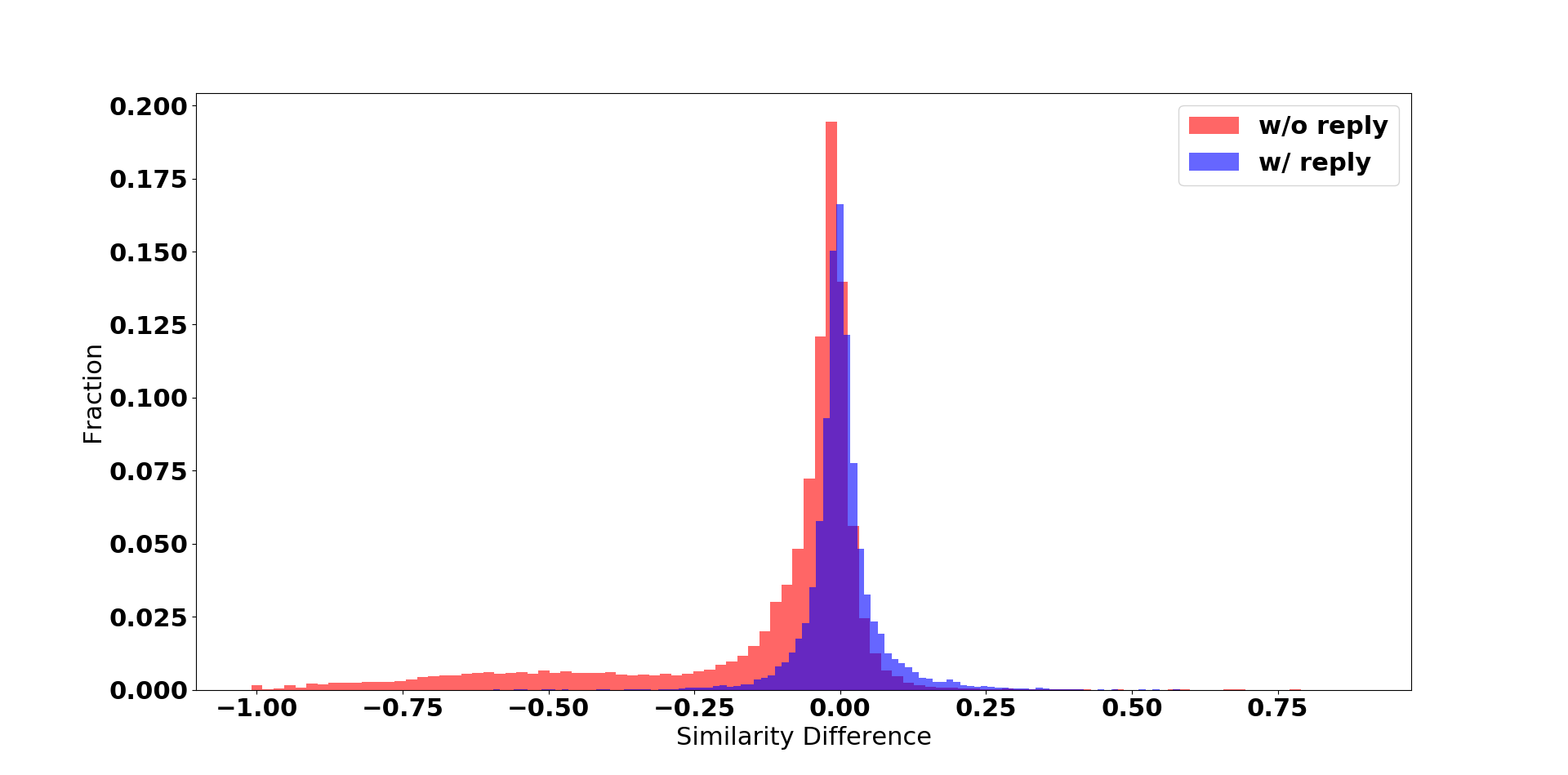}
        \caption{TFIDF}
    \end{subfigure}
    \begin{subfigure}[b]{0.48\textwidth}
        \includegraphics[width=\textwidth]{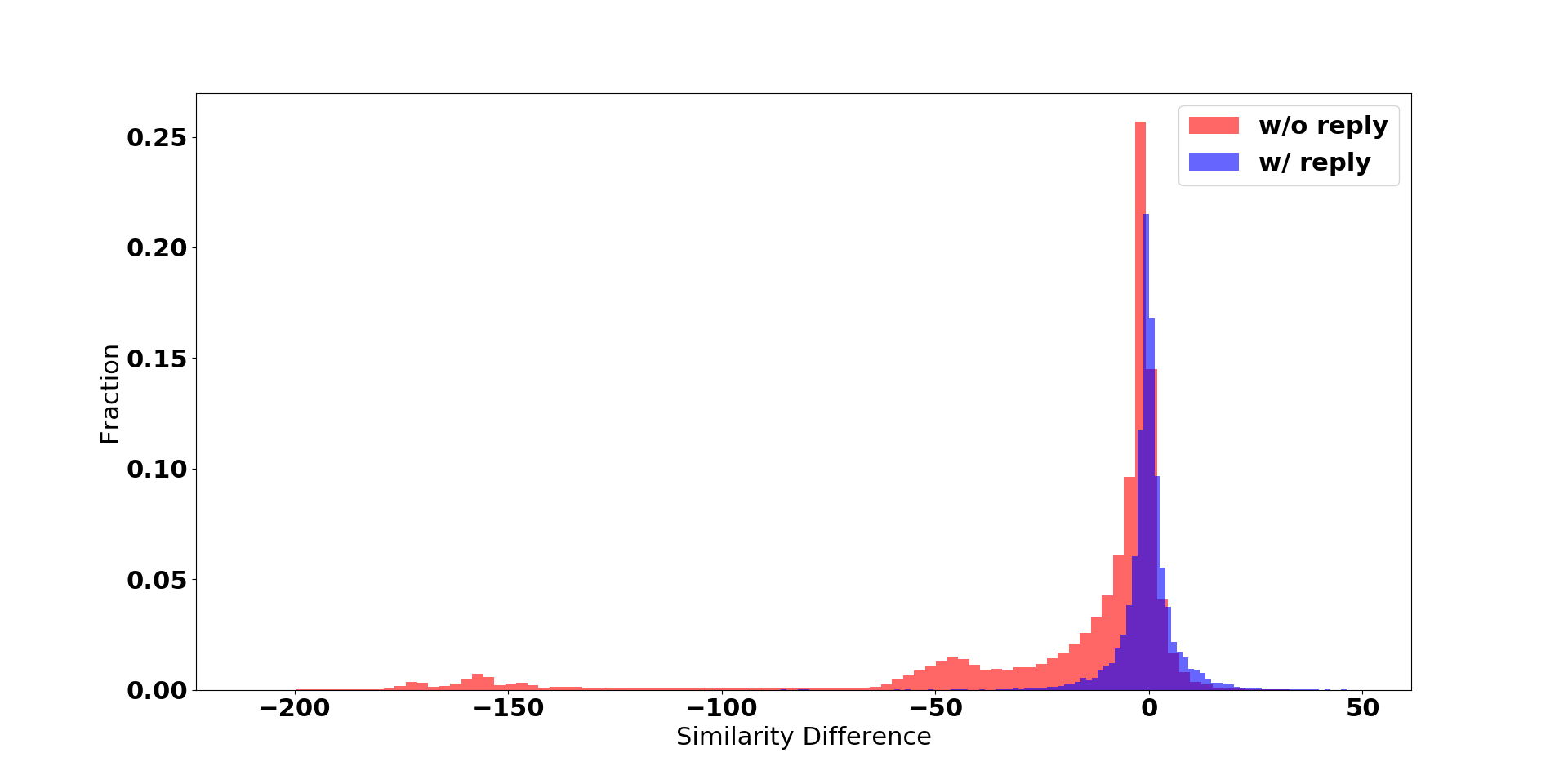}
        \caption{CNN}       
    \end{subfigure}
    \caption{Histogram of the contrastive term (the difference between the similarity of the email and positive user representation and that of the email and negative user representation for each class on the test set. History length here is $10$. The top plot shows the results for the TFIDF content representation, while the bottom plot is the CNN counterpart. Best viewed in color.}
\label{fig:sim_diff}
\end{figure}

Table \ref{tab:sim} shows the results from the ablation study for using the \texttt{Pos+Neg} user representations. We can see that the similarity features are the decisive factors here. Without them, the performance of the traditional TFIDF representation is unable to compete with the more advanced neural embeddings based representations. However, the introduction of the similarity terms change the situation entirely and improve the performance of both representations, which leads to the competitive performance of the TFIDF based representations. To examine if the classifier itself play a role here, we also experiment with the GBDT classifier trained on top of the learned CNN representations. We obtain an averaged test AUROC of $0.7767$ which is comparable to $0.7764$ from the MLP and eliminates the possible impact from different classifiers.

From the feature importance analysis of the TFIDF-GBDT model, we see that all three similarity terms introduced in Section \ref{sec:simfeat} appear among the top important features, where the difference between the similarity with the positive and negative histories is the most influential one. Figure \ref{fig:sim_diff} shows the histograms of this feature for the replied/non-replied emails in the test set with both TFIDF and CNN representations respectively. From the histograms, we can see that the values of majority of the emails the user replied to are narrowly distributed with a positive mode, while the non-replied class has a heavy long tail to the left with a negative mode. It is evident that there is noticeable difference between the two classes even from this feature only. This also explains the remarkable benefit coming from incorporating the negative history to represent users shown in Table \ref{tab:meta}. Note that the similarity term for the TFIDF representations compute the weighted number of matching word grams in both the incoming and historical emails. This also gives a way to interpret the classification through examining positive and negative correlated word tokens.

\begin{table}[h]
\centering \ra{1.2}
\begin{tabular}{@{}rr@{}} \toprule
\textbf{Model} & \textbf{Test AUROC} \\ \midrule
TFIDF-GBDT & 0.7773 \\
CNN-MLP & 0.7764 \\
Ensemble & \textbf{0.7880} \\ \bottomrule
\end{tabular}
\caption{Test AUROC results of the ensemble method. Here, the ensemble result is computed from simple average of the predicted probabilities of the TFIDF-GBDT and CNN-MLP models where the \texttt{Pos+Neg} user representation and history length $10$ are used. The best performance is shown in bold.}
\label{tab:ensemble}
\end{table}

\subsubsection{Ensemble} 
Considering the competitive performance from the TFIDF representations, it is tempting to think if the CNN models are also just picking up the matching word grams. To verify this, we look at the simplest ensemble model, where the final predicted probability is the average of those from both the TFIDF-GBDT and the CNN-MLP models. The results are shown in Table \ref{tab:ensemble} . We can see that the simple ensemble model gives additional performance boost, resulting a $1.4\%$ AUROC improvement. This also suggests that the traditional word token based and the neural embedding based representations are complementary to each other, and can be beneficial if used together.

\begin{table}[h]
\centering \ra{1.3}
\begin{tabular}{@{}rr@{}}\toprule
\textbf{Model} & \textbf{Test AUROC} \\ \midrule
 User Embedding-MLP ($d=50$) \cite{ncf} & 0.7093 \\
 User Embedding-MLP ($d=300$) \cite{ncf} & 0.7131 \\ \midrule
 
 \texttt{Pos}-CNN-MLP (ours) & 0.7457 \\ \midrule
 
 \texttt{Pos+Neg}-CNN-LWA \cite{vartak} & 0.7511 \\ 
 \texttt{Pos+Neg}-CNN-NLBA \cite{vartak} & 0.7570 \\ \midrule
 
 \texttt{Pos+Neg}-CNN-MLP w/o similarity (ours) & 0.7541 \\
 \texttt{Received}-CNN-MLP (ours) & 0.7592 \\
 \texttt{Pos+Neg}-CNN-LR (ours) & 0.7598 \\ 
 \texttt{Received}-TFIDF-GBDT (ours) & 0.7622 \\
 \texttt{Pos+Neg}-CNN-MLP (ours) & \textbf{0.7764} \\ 
 \texttt{Pos+Neg}-TFIDF-GBDT (ours) & \textbf{0.7773} \\ \bottomrule
\end{tabular}
\caption{Comparison with existing deep recommendation methods. Here, ``user embedding" is learning an explicit embedding for each user from the one-hot indicator vector. The user representation with history length up to 10 is considered for the rest models. Here, the aggregation function used is the dot-based attention \eqref{eqn:dot}.}
\label{tab:comparison}
\end{table}

\subsubsection{Comparison with Existing Methods}
\label{sec:compexist}
We compare our approach against the deep recommendation baselines. For fair comparison, we use the same convolutional network architecture to extract email representations. Except for the NCF method \cite{ncf} which learns explicit user embeddings, we use a user history length up to $10$. To directly compare with the two adaptation methods, we consider both the LR and MLP classifiers on top of the email and user representations, where the MLP model contains a single hidden layer of size $128$. We do not observe improved validation performance with additional hidden layers.

Table \ref{tab:comparison} shows the averaged test AUROC results (the standard deviations are smaller than $0.003$). The adapted NCF method \cite{ncf} performs worst among all models, which is somewhat expected since it learns explicit user representation via embedding matrix on top of the one-hot indicator user vectors. Even though it can still summarize the users' reply preferences through the interaction with the content of incoming emails during training, the learned user representations are limited to only the users present in the training set and the response patterns during the training time period. The positive user representations \cite{dkn,interest}, on the other hand, can directly summarize user preferences without solely relying on the supervised signal during training. Thus it shows performance improvement\footnote{The number reported in Table \ref{tab:comparison} for the ``\texttt{Pos}-CNN-MLP" model actually comes from the MLP trained with the additional similarity features incorporated which were not considered in \cite{dkn,interest}. This gives better performance than the original methods, similar to the situation shown in Table \ref{tab:sim}.} over the NCF method, and can learn time-insensitive interactions between user and emails.

The adaptation methods \cite{vartak} gives the best performance among all the baselines, due to the utilization of both positive and negative histories and the adaptation strategies. But our approaches can still present improvement even without attending to the best-performing \texttt{Pos+Neg} user representation. With the \texttt{Pos+Neg} user representation, we are able to achieve a $2.6\%$ AUROC improvement over the best baseline result. We attribute the superiority of our proposed framework to its $simple but$
effective utilization of the representations. In the adaptation methods \cite{vartak}, the deep classifier is trained on the email representation where the bias of each layer is a function of the user representation. On the other hand, ours directly train the classifier on top of the concatenation of both representations and the similarity between them. Without the similarity, the performance of the CNN-MLP model is slightly worse than that of the NLBA model. As discussed in Section \ref{sec:sim}, the similarity features not only provide enough predictive power through the contrast with the class ``prototypes", but also free the classifier to attend more to the local details and interaction patterns instead of the global resemblance.

\subsection{Impact Analysis}
In this section, we examine how different choices of the various components impact the performance of our proposed framework. 

\subsubsection{Impact of History Length}
\label{sec:length}
We study the impact of different history lengths on the prediction performance, which is illustrated in Table \ref{tab:history}. We can see that the performance is gradually improved as more historical emails are included. But the marginal improvement is limited, especially for the \texttt{Pos+Neg} user representation. The only exception is the \texttt{Pos}-CNN-MLP model, where we can still observe noticeable improvement beyond history length $10$.
 
Another question we try to answer here is whether more historical emails considered in the \texttt{Pos+Neg} user representation contribute to the better performance, since it uses twice the number of historical emails used by the \texttt{Pos} representation. The answer is no. Adding more positive emails can not compensate the benefit of incorporating negative samples. Indeed, with up to 6 historical emails (3 positive and 3 negative), it already gives $2.5\%$ and $5.5\%$ relative AUROC improvement over using 20 positive emails for the CNN-MLP and TFIDF-GBDT models respectively. This shows that the \texttt{Pos+Neg} representation is actually more data efficient. Another thing worth pointing out is that the CNN-MLP model is the clear winner in the limited information scenarios, where either only positive history or limited history length is considered.

\begin{table}[h]
\centering \ra{1.2}
\begin{tabular}{@{}rrrrrr@{}}\toprule
\multirow{2}{*}{\textbf{User}} & \multirow{2}{*}{\textbf{Model}} & \multicolumn{4}{c}{\textbf{History Length}} \\ \cmidrule{3-6}
& & \textbf{3} & \textbf{5} & \textbf{10} & \textbf{20} \\ \midrule
\multirow{2}{*}{\texttt{Pos}} & TFIDF-GBDT & 0.7205 & 0.7233 & 0.7248 & 0.7277 \\
& CNN-MLP & 0.7383 & 0.7441 & 0.7457 & 0.7515 \\ \midrule
\multirow{2}{*}{\texttt{Pos+Neg}} & TFIDF-GBDT & 0.7675 & 0.7719 & 0.7773 & 0.7796 \\
& CNN-MLP & 0.7703 & 0.7735 & 0.7764 & 0.7767 \\ \bottomrule
\end{tabular}
\caption{Comparison of test AUROC results under different lengths of user histories. Here, the top section shows results where users are represented using the \texttt{Pos} representation, while the bottom section shows results where users are represented using the \texttt{Pos+Neg} representation.}
\label{tab:history}
\end{table}

\subsubsection{Impact of Attention} 
We investigate how different aggregation functions affect the deep learning performance, which is shown in Table \ref{tab:attention}. Unlike \cite{attention}, we do not observe significant difference between the performance of the attention-based user representation and that of the global user representation. Even though the attention model does provide email-specific user representation which manifest a marginal improvement, the resulting attention weights are not far from the globally learned ones. In fact, they are all minutely deviated from $1/h$, where $h$ is the history length considered. This also sheds some light on why the simple average of the history using the TFIDF representation performs well. For the concat-based attention model \eqref{eqn:attention} and \eqref{eqn:concat}, we are unable to get reasonable results. This might be similar to the situation shown in Table \ref{tab:sim} where the inner product term plays a crucial role.

We also try to interpret the predictions from the attention weights. We select the largest attention weights and retrieve the corresponding historical positive emails. Examining those emails and comparing with the new incoming email, we find that they all share some common words \footnote{Content is not disclosed due to the requirement of the License Agreements of the Avocado dataset.} indicative of the possible reply action. The principal reason behind the failure of complicated models such as concat-based attention or a deeper neural architecture is the predominant lexical nature of the problem.

Unlike sentiment analysis \cite{kim} or news recommendation \cite{dkn}, the length of the emails are much longer. Rather than extracting high order semantics, it is more important for the model to show the contrasts between the current email and the replied/non-replied aggregated contents. This is evident also through the dominance of the similarity features.

\begin{table}[h]
\centering \ra{1.2}
\begin{tabular}{@{}rr@{}}\toprule
\textbf{Aggregation} & \textbf{Test AUROC} \\ \midrule
Learned global weights & 0.7745 \\
Attention (dot) & 0.7764 \\
Attention (concat) & 0.7209 \\ \bottomrule
\end{tabular}
\caption{Comparison of test results under different aggregation functions for the CNN-MLP model where users are represented using the \texttt{Pos+Neg} representation with history length up to 10.}
\label{tab:attention}
\end{table}

\subsubsection{Impact of Initialization of Word Embedding} 
The input to the convolutional neural networks requires an embedding look-up table for word tokens before the convolution using filters. In order to investigate the impact of initializations, we experiment with many different techniques including random initialization, and many pre-trained embeddings from word2vec \cite{word2vec}, GloVe \cite{glove} to fasttext \cite{fasttext}. We also train the embeddings directly on the Avocado training and validation datasets, considering the different structures of email corpus from Wikipedia or GoogleNews (which are used for the pre-trained embeddings). We consider both word2vec and fasttext with the skip-gram model and a window size of $5$. The best performance is obtained with static embeddings trained on the Avocado corpus (Details in Appendix).

\section{Related Works}
\label{sec:relatedwork}
Email prioritization has previously been investigated. In \cite{gmail}, a per-user logistic regression model is learned to predict the probability that a user will perform certain action on the incoming email. This is similar to the problem considered in our work, where the prediction is from the recipient's perspective. This is considerably complicated than the problem considered in \cite{yang} especially for the email that was sent to multiple users. The authors considered different categories of features including social, thread, label and content features. The content features attempt to identify headers and word terms that are highly correlated with the recipient acting (or not) on the email. Since the ``correlation" is learned offline, it must be re-trained frequently. This can be considered as a simplified version of our proposed similarity-augmented framework. However, it is a local model where the terms are directly identified for each user, while ours is a global model where we try to discover the relationship between the email and user representations. In this way, our model does not need to be re-trained frequently.

Personalized email prioritization through collaborative filtering has also been studied for broadcast emails \cite{broadcast}, where the prediction is based on the feedback from a small subset of recipients of the same email. Despite showing great performance, the problem scope considered is rather limited. It can only be applied to emails sent to large number of recipients and can not perform the prediction on delivery time since it needs to wait for responses from some subset of the recipients. Therefore, it can not be applied to the general prediction problem including predictions for emails sent to limited number of recipients and time-sensitive situations.

Modeling the relation between user and item has been well studied in the literature on recommendation systems \cite{tang,yu2016,benton,song}. Although these approaches reported improved performance on specific tasks, they all rely on explicitly learning the user representations from the one-hot indicator vector. This faces the acute problem of extending beyond users in the training set. Even for these users, it is unable to adapt to their changing response patterns thus still needs to be re-trained frequently. On the other hand, our content-based user representation can automatically adapt to the users' recent response patterns. What's more, the globally learned interaction between email and user representations can be directly applied to new users once they start to receive emails without re-training on those users.

A more scalable approach, as mentioned above, is to use the contents of the items for user representation which has recently been extensively studied in the recommendation community \cite{content,blei,dkn,interest,vartak} and references therein. The most popular user representation is generated using the contents of items that the user has positively interacted with, \cite{blei,dkn,interest} for example. However, as shown in Section \ref{sec:experiment}, representing users using positive history only does not perform well (even worse than using the received history). This may be because of the different characteristics of problems considered: the reply actions studied in our work is considerably rare compared to the click/read actions considered in traditional recommendation problems. The limited contents of the replied emails and the lack of contrastive information from both positive and negative histories make it difficult to perform well.

Our approach, especially with users represented using both positive and negative histories, can be seen as an variant of the multi-task learning framework \cite{multitask} applied to the prototypical networks \cite{prototype} with shared architecture. The prototypical network is proposed for few-shot learning, where each class is represented by a prototype representation and the classification is performed by computing distances to all the prototypes. In our scenario, action prediction for each user corresponds to one \textit{task}, and the \textit{prototypes} are the aggregated user positive and negative representations. The shared architecture provides knowledge transfer among users.

\section{Conclusion and Discussions}
\label{sec:conclusion}
In this work, we propose the recommendation framework for the content-based email triage action prediction problem and conduct in-depth analysis of various email and user representations. It addresses several major challenges in the prediction problem. First, it is a content-based representation method which does not rely on explicit user embedding, thus can be easily adapted to unseen users. Historically received emails provide a natural way to obtain personalization and adaptation. We show that the popular positive-history based user representation fails to compete even with the received history without considering any ``supervised" information. On the other hand, incorporating class-specific user representations from both positive and negative user histories gives best performance. Second, we introduce additional similarity features besides the emails and user representations which are able to achieve better performance compared to the recently proposed state-of-the-art deep recommendation methods. In addition, it also helps the traditional bag-of-words content representation to compete favorably with the neural embedding based methods. More importantly, we show that the best performance is achieved by combining the predictions from both the traditional and neural embedding based models, which highlights the different aspects conveyed from divergent content representations. While this work focused on the email contents and reply prediction as a working example, the framework developed here and the insight regarding different representations can be readily applied to other online recommendation problems possibly with different features.

\section{Acknowledgement}

Sudipto Mukherjee would like to thank Microsoft Corporation for providing the internship opportunity in Summer 2018 during which time this work was completed.

\bibliography{Ref}

\begin{thebibliography}{10}

\bibitem{tensorflow}
Martin Abadi, Paul Barham, Jianmin Chen, Zhifeng Chen, Andy Davis, Jeffery
  Dean, Matthieu Devin, Sanjay Ghemawat, Geoffrey Irving, Michael Isard,
  Manjunath Kudlur, Josh Levenberg, Rajat Monga, Sherry Moore, Derek~G. Murray,
  Benoit Steiner, Paul Tucker, Vijay Vasudevan, Pete Warden, Martin Wicke, Yuan
  Yu, and Xiaoqiang Zheng.
\newblock Tensor{F}low: a system for large-scale machine learning.
\newblock In {\em Proceedings of OSDI'16}, 2016.

\bibitem{gmail}
Douglas Aberdeen, Ondrej Pacovksy, and Andrew Slater.
\newblock The learning behind gmail priority inbox.
\newblock In {\em LCCC: NIPS 2010 Workshop}, 2010.

\bibitem{benton}
Adrian Benton, Raman Arora, and Mark Dredze.
\newblock Learning multiview embeddings of twitter users.
\newblock In {\em Proceedings of the 54th Annual Meeting of the Association for
  Computational Linguistics (Volume 2: Short Papers)}, 2016.

\bibitem{randomsearch}
James Bergstra and Yoshua Bengio.
\newblock Random search for hyper-parameter optimization.
\newblock {\em Journal of Machine Learning Research}, 13:281--305, 2012.

\bibitem{fasttext}
Piotr Bojanowski, Edouard Grave, Armand Joulin, and Tomas Mikolov.
\newblock Enriching word vectors with subword information.
\newblock {\em Transactions of the Association for Computational Linguistics},
  5:135--146, 2017.

\bibitem{multitask}
Rich Caruana.
\newblock Multitask learning.
\newblock {\em Machine Learning}, 28:41--75, 1997.

\bibitem{ncf}
Xiangnan He, Lizi Liao, Hanwang Zhang, Liqiang Nie, Xia Hu, and Tat-Seng Chua.
\newblock Neural collaborative filtering.
\newblock In {\em Proceedings of the 26th International Conference on World
  Wide Web (WWW)}, 2017.

\bibitem{lightgbm}
Guolin Ke, Qi~Meng, Thomas Finley, Taifeng Wang, Wei Chen, Weidong Ma, Qiwei
  Ye, and Tie-Yan Liu.
\newblock Light{GBM}: a highly efficient gradient boosting decision tree.
\newblock In {\em Advances in Neural Information Processing Systems (NIPS)},
  2017.

\bibitem{kim}
Yoon Kim.
\newblock Convolutional neural networks for sentence classification.
\newblock {\em Proceedings of the 2014 Conference on Empirical Methods in
  Natural Language Processing (EMNLP)}, 2014.

\bibitem{adam}
Diederik~P. Kingma and Jimmy Ba.
\newblock Adam: a method for stochastic optimization.
\newblock In {\em Proceedings of the 3rd International Conference on Learning
  Representations (ICLR)}, 2015.

\bibitem{content}
Pasquale Lops, Marco~de Gemmis, and Giovanni Semeraro.
\newblock {\em Content-based recommender systems: state of the art and trends}.
\newblock Springer, 2011.

\bibitem{attention}
Minh-Thang Luong, Hieu Pham, and Christopher~D. Manning.
\newblock Effective approaches to attention-based neural machine translation.
\newblock In {\em Proceedings of the 2015 Conference on Empirical Methods in
  Natural Language Processing (EMNLP)}, 2015.

\bibitem{word2vec}
Tomas Mikolov, Ilya Sutskever, Kai Chen, Greg~S Corrado, and Jeff Dean.
\newblock Distributed representations of words and phrases and their
  compositionality.
\newblock In {\em Advances in neural information processing systems (NIPS)},
  2013.

\bibitem{avocado}
Douglas Oard, William Webber, David Kirsch, and Sergey Golitsynskiy.
\newblock {\em Avocado Research Email Collection LDC2015T03}.
\newblock Philadelphia: Linguistic Data Consortium, 2015.

\bibitem{glove}
Jeffrey Pennington, Richard Socher, and Christopher Manning.
\newblock Glove: Global vectors for word representation.
\newblock In {\em Proceedings of the 2014 conference on empirical methods in
  natural language processing (EMNLP)}, 2014.

\bibitem{bow}
Gerard Salton and Christopher Buckley.
\newblock Term-weighting approaches in automatic text retrieval.
\newblock {\em Information Processing \& Management}, 24:513--523, 1988.

\bibitem{prototype}
Jake Snell, Kevin Swersky, and Richard Zemel.
\newblock Prototypical networks for few-shot learning.
\newblock In {\em Advances in Neural Information Processing Systems (NIPS)},
  2017.

\bibitem{song}
Yan Song and Chia-Jung Lee.
\newblock Learning user embeddings from emails.
\newblock In {\em Proceedings of the 15th Conference of the European Chapter of
  the Association for Computational Linguistics: Volume 2, Short Papers}, 2017.

\bibitem{tang}
Duyu Tang, Bing Qin, and Ting Liu.
\newblock Learning semantic representations of users and products for document
  level sentiment classification.
\newblock In {\em Proceedings of the 53rd Annual Meeting of the Association for
  Computational Linguistics and the 7th International Joint Conference on
  Natural Language Processing (Volume 1: Long Papers)}, 2015.

\bibitem{vartak}
Manasi Vartak, Arvind Thiagarajan, Conrado Miranda, Jeshua Bratman, and Hugo
  Larochelle.
\newblock A meta-learning perspective on cold-start recommendations for items.
\newblock In {\em Advances in Neural Information Processing Systems (NIPS)},
  2017.

\bibitem{broadcast}
Beidou Wang, Martin Ester, Jiajun Bu, Yu~Zhu, Ziyu Guan, and Deng Cai.
\newblock Which to view: personalized prioritization for broadcast emails.
\newblock In {\em Proceedings of the 25th International Conference on World
  Wide Web (WWW)}, 2016.

\bibitem{blei}
Chong Wang and David~M Blei.
\newblock Collaborative topic modeling for recommending scientific articles.
\newblock In {\em Proceedings of the 17th ACM SIGKDD international conference
  on Knowledge discovery and data mining}, pages 448--456. ACM, 2011.

\bibitem{dkn}
Hongwei Wang, Fuzheng Zhang, Xing Xie, and Minyi Guo.
\newblock D{KN}: deep knowledge-aware network for news recommendation.
\newblock In {\em Proceedings of the 27th International Conference on World
  Wide Web (WWW)}, 2018.

\bibitem{yang}
Liu Yang, Susan~T Dumais, Paul~N Bennett, and Ahmed~Hassan Awadallah.
\newblock Characterizing and predicting enterprise email reply behavior.
\newblock In {\em Proceedings of the 40th International ACM SIGIR Conference on
  Research and Development in Information Retrieval}, 2017.

\bibitem{yu2016}
Yang Yu, Xiaojun Wan, and Xinjie Zhou.
\newblock User embedding for scholarly microblog recommendation.
\newblock In {\em Proceedings of the 54th Annual Meeting of the Association for
  Computational Linguistics (Volume 2: Short Papers)}, 2016.

\bibitem{interest}
Guorui Zhou, Chengru Song, Xiaoqiang Zhu, Ying Fan, Han Zhu, Xiao Ma, Yanghui
  Yan, Junqi Jin, Han Li, and Kun Gai.
\newblock Deep interest network for click-through rate prediction.
\newblock In {\em Proceedings of the 24th ACM SIGKDD Conference on Knowledge
  Discovery and Data Mining (KDD)}, 2018.

\end{thebibliography}
\bibliographystyle{plain}

\section*{Appendix}
\textbf{Reproducibility}.
Keeping reproducibilty as our prime focus, we have used publicly available email collection for our experiments instead of proprietary data. All software tools used in this work are open-source. The main text contained details of data processing steps and algorithm parameters where possible. In addition, we provide all the hyper-parameter details in this section that can help reproduce the results.

We show the parameter settings for different models considered in our framework. Random search \cite{randomsearch} is used to find the best parameter setting. 

For the logistic regression model, the parameters considered are shown in Table \ref{tab:lr}. We mainly consider the regularization and positive weights to counter the highly imbalanced issue. The best parameters found for the TFIDF representation are $1.0$ for regularization and $1.0$ for positive weight. For the Embed representation, they are $1.0$ for regularization and ``balanced'' for positive weight.

\begin{table}[h!]
\centering \ra{1.2}
\begin{tabular}{ @{}rr@{} } \toprule 
\textbf{Hyper-parameter} &  \textbf{Value} \\ \midrule
Inverse Reg. Coeff. & 0.01, 0.1, 1, 10, 100 \\  
Positive weight &  1, 5, 10, 15, 'balanced' \\  \bottomrule
\end{tabular}
\caption{Various LR hyper-parameters considered.}
\label{tab:lr}
\end{table}

For the GBDT model, the parameters considered are shown in Table \ref{tab:gbdt}. We mainly concern about the number of trees, maximum depth, learning rate and positive weight. The best result for the TFIDF representation is from $500$ trees with maximum depth $5$ learned with learning rate $0.1$ and positive weight $5.0$. For the Embed representation, the best result is from $300$ trees with maximum depth $5$ learned with learning rate $0.1$ and positive weight $5.0$.

For the CNN model, Table \ref{tab:tune} shows all the possible parameter values considered during random search and the best parameter settings are shown in Table \ref{tab:best} for using the \texttt{Pos} user representation and the \texttt{Pos+Neg} user representation respectively.

\begin{table}[h]
\centering \ra{1.2}
\begin{tabular}{ @{}rr@{} } \toprule 
\textbf{Hyper-parameter} &  \textbf{Value} \\ \midrule
\# Iterations & 50, 100, 200, 300, 500, 800 \\ 
Max. Depth & 2, 3, 4, 5 \\ 
\# Leaves  & $2^{\textrm{Max. Depth}} - 2$ \\ 
Positive weight &  5, 10, 15, 20, 'balanced' \\ 
Learning Rate & 0.01, 0.1, 1.0 \\ \bottomrule
\end{tabular}
\caption{Various GBDT hyper-parameters considered.}
\label{tab:gbdt}
\end{table}

\begin{table}[h!]
\centering \ra{1.2}
\begin{tabular}{ @{}rr@{} } \toprule 
\textbf{Hyper-parameter} &  \textbf{Values} \\ \midrule
Filter Sizes & 1, 2, 3, 4, 5 \\ 
\# Filters &  64, 128, 256 \\ 
Sequence Length & 75, 100, 150 \\ 
Embed Dim & 50, 100, 300 \\
Aggregate $\mathcal{G}$ & linear transform, Attention \\ 
\# Hidden Units in MLP & 128 \\ 
Batch size & 32, 64, 128, 256 \\ 
Learning Rate & 0.001, 0005, 0.0001 \\ 
Dropout & 0.5, 1.0 \\ 
Positive weight & 5.0, 10.0, 15.0, 20.0 \\ \bottomrule
\end{tabular}
\caption{Various CNN hyper-parameters considered.}
\label{tab:tune}
\end{table}

\begin{table}[h!]
\centering \ra{1.2}
\begin{tabular}{ @{}rrr@{} } \toprule 
\multirow{2}{*}{\textbf{Hyper-parameter}} & \multicolumn{2}{c}{\textbf{User Representation}} \\ \cmidrule{2-3}
 & \texttt{Pos} & \texttt{Pos+Neg} \\ \midrule
Filter Sizes & 1, 2 & 1,2,3 \\ 
\# Filters & 64, 64 & 256,128,64 \\ 
Sequence Length & 75 & 150 \\ 
Embed Dim & 50 & 100 \\
Aggregate $\mathcal{G}$ & Dot-Attention & Dot-Attention \\ 
\# Hidden Units in MLP & 128 & 128 \\ 
Batch size & 32 & 128 \\ 
Learning Rate & 0.0005 & 0.0005 \\ 
Dropout & 0.5 & 0.5 \\ 
Positive weight & 15.0 & 10.0 \\ \bottomrule
\end{tabular}
\caption{Best CNN hyper-parameters found.}
\label{tab:best}
\end{table}

\begin{table}[h!]
\centering \ra{1.2}
\begin{tabular}{ @{}rrr@{} } \toprule
\textbf{Embed Type} & \textbf{Dim} & \textbf{Val AUROC} \\ \midrule
Avo-fasttext & 50 & \textbf{ 0.7779 $\pm$ 0.0010} \\ 
Avo-word2vec & 50 &  0.7764 $\pm$ 0.0011 \\ 
GloVe & 50 &  0.7676 $\pm$ 0.0008 \\ 
GloVe & 300 &  0.7718 $\pm$ 0.0011 \\ 
word2vec & 300 &  0.7713 $\pm$ 0.0013 \\ 
fasttext & 300 &  0.7717 $\pm$ 0.0010 \\ 
random & 50 &  0.7537 $\pm$ 0.0022 \\ \bottomrule
\end{tabular}
\caption{Comparison of the validation AUROC results of the CNN-MLP model using different word embeddings and the \texttt{Pos} user representation. Here, "Avo-" prefix indicates word embeddings learned from the Avocado dataset directly, where embedding type without any prefix is the pre-trained embeddings. The best performance is shown in bold.}
\label{tab:embed}
\end{table}

\noindent
\textbf{Impact of Word Embeddings}.
The validation results are shown in Table \ref{tab:embed} for the \texttt{Pos} user representations. Similar trends are also observed for the \texttt{Pos+Neg} ones. We can see that random initialization gives the worst performance, showing the importance of proper initialization here. The pre-trained word embeddings on large text corpus give considerable improvement over the random counterpart. However, the best performance comes from the self-trained embeddings on the Avocado dataset directly, despite the limited embedding dimension and much smaller training corpus. We also experiment with larger dimensions for the self-trained embeddings, but do not see noticeable difference. In addition, we do not observe any improvement from fine-tuning the word embeddings through joint training.

\end{document}